\pdfoutput=1
\documentclass[twocolumn,preprintnumbers,amsmath,amssymb,aps,prl,floatfix,
superscriptaddress]{revtex4}
\usepackage{graphicx}
\usepackage{dcolumn}
\usepackage{bm}
\usepackage{amssymb,amsmath}

\begin{document}

\title{Interacting Faults in California and Hindu Kush}

\author{Callum Muir}
\affiliation{Center for Nonlinear Science, University of North Texas, P.O. Box 311427, 
Denton, Texas 76203-1427, USA }

\author{Jordan Cortez}
\affiliation{Earth and Planetary Sciences, University of California Riverside, 
Riverside, California 92521, USA}

\author{Paolo Grigolini}
\affiliation{Center for Nonlinear Science, University of North Texas, P.O. Box 311427, 
Denton, Texas 76203-1427, USA }

\begin{abstract}
We study seismic fluctuations  in California and Hindu Kush using Diffusion Entropy Analysis (DEA), a technique  designed to detect the action of crucial events in time series generated by complex 
dynamical systems. The time distance between two consecutive crucial events is described by an inverse power law distribution density with a power index $\mu$ close to the value $\mu = 2$, corresponding to an ideal $1/f$ noise. DEA was used in the recent past to study neurophysiological processes that in the healthy condition are found to generate  $1/f$ noise and $\mu$ close to $2$. In this paper we find that the seismic fluctuations in both California and Hindu-Kush of extended areas implying the action of many faults, $\mu \approx  2.1$, while in regions involving the action of only one fault, or of a very small number of faults, $\mu \approx 2.4$. This observation led us the conjectures that the  seismic criticality is due to the interaction of many faults. To support this conjecture we adopted a dynamical model for fault dynamics proposed by Brown and Tosatti and we have extended it to describe the interaction between many faults. The DEA applied to surrogate sequences generated by this dynamical model, yields $\mu = 2.1$ for a single fault and $\mu = 2,4$ for many interacting faults, in a good agreement with the observation of real seismic fluctuations. This result supports our conjecture and suggests interesting applications to neurophysiological and sociological processes.

\end{abstract}
\maketitle

\section{Introduction}
Modeling earthquakes is one of the most important issues of the modern field of complexity.  The popular theory of Self-Organized Criticality (SOC) \cite{soc} originally created to explain $1/f$ noise led the investigators to discover the connection between criticality and avalanches, a property of interest more general than the merely geophysical, as proved by the fact that Begg and Plenz in 2003 found avalanches  in neocortical circuits \cite{plenz}. The paper of Ref. \cite{plenz} led the researchers in the field of complexity to the conjecture that avalanches are a signature of criticality,  a general property of systems of cooperating units,  ranging from earthquakes \cite{bakonquakes} to the brain \cite{chialvo}. 

Another significant example of the general interest for modeling earthquakes is given by the work of Sornette and Helmstetter \cite{sornette2,agnes,subordination}.
 These authors proposed a model where each shock is a generator of another shock \cite{sornette2} and two consecutive shocks may have different mothers. As a consequence the distribution density of inter-event times gets a special form, reminiscent of a Weibull function, that has been adopted for applications to the financial market \cite{financialmarket}. 
 
 The modeling of earthquakes depends to a great extent by the results of the analysis of seismic fluctuations, namely, the technique adopted to analyze time series. The recent paper by
 Jim\'{e}nez \cite{abigail} on the adoption of the Hurst analysis \cite{hurst} and of Diffusion Entropy Analysis (DEA) \cite{scafetta}. The Hurst analysis has been used in a recent paper 
 \cite{hindukush}.
 
As far as the Hurst method is concerned, it is convenient to notice, as pointed out in the 1994 paper by Mannella \emph{et al} \cite{mannella}, that the Hurst method is interpreted as a realization of the Fractional Brownian Motion (FBM)\cite{mandelbrot}, which is based on correlated Gaussian processes, but in some
cases the meaning 
of the Hurst coefficient $H$
may not coincide with
the scaling of  the diffusion
process evaluated with 
the second moment 
technique.

The DEA, on the contrary, was originally created \cite{scafetta} for the purpose of evaluating in a rigorous way scaling also in the cases where the diffusion process departs from the condition of finite second moments, but in principle \cite{scafettaandbruce} it cannot establish if the anomalous scaling coefficient is due to FBM , which keeps finite the second moments, or to a process generating at the same time deviation from the ordinary scaling and long tails with diverging second moments. 
 
 It has to pointed out that there exists a recent version of DEA \cite{garland} that filters the anomalous contributions generated by FBM, if they occur.  This updated version of DEA was designed to detect crucial events. The crucial events are bursts of randomness characterized by the following mai properties: (a)  The time distance between two consecutive bursts of randomness is described by a waiting time distribution density with the form 
 \begin{equation} \label{manneville}
 \psi(\tau) = (\mu-1) \frac{T^{\mu-1}}{(\tau + T)^{\mu}},
  \end{equation}
  with $\mu <3$; (b) The occurrence of these events fits
  the renewal condition, namely, the occurrence of an event generates a new event at a time $\tau$ later with no correlation with the earlier time distances.

 In this paper we analyze seismic fluctuations with this updated version of DEA \cite{garland} and find, in a apparent conflict, with Jimenez that when our analysis is done on wide areas involving many faults, the inverse power index $\mu$ is very close to $\mu = 2$, in accordance with the earlier results of the paper of Ref. \cite{mega}, yielding $\mu = 2.06$. When only one fault is considered we find much larger values of $\mu$ such as $\mu = 2.5$. 
 
 The main aim of this paper is to prove that the criticality condition in the case of seismic fluctuations is due to the interaction between different faults. The  big change from high values of $\mu$ in the case of single faults to values of $\mu$ much closer to the ideal value $\mu = 2$ are a consequence of the interaction between faults. In this sense the whole geophysical underground can be perceived as a complex system of interacting units, each fault being a unit of the system. This is in line with the original work of Bak \cite{soc,nature} with emphasis though on temporal complexity \cite{korosh}, namely the time distance between two consecutive critical fluctuations,  rather than on the size of fluctuations. Temporal complexity leads to the discovery that $1/f$ noise is generated by crucial events, the noise spectrum being $S(f) \propto 1/f^{\beta})$, with $\beta = 3 -\mu$ \cite{mirko}.  On the basis of this approach to $1/f$ noise, the authors of Ref. \cite{brain}  reached the conclusion that the brain of healthy awake individuals, with $\mu = 2$ is a source of ideal $1/f$ noise. Therefore we plan to prove that the geophysical complexity is the same as that of the human brain.
 
 The outline of the paper  is as follows. In Section \ref{observation} we use the updated DEA of Ref. \cite{garland} to analyze California earthquakes and we find that  $\mu$ ranges from $\mu = 2.5$ for single fault to $\mu = 2.1$ for areas involving many faults. In Section \ref{tosatti} we run sequences according to the theory of Ref. \cite{erio} and we analyze them using again he updated DEA of Ref. \cite{garland}. In Section \ref{interacting} we study the interaction between many faults.  
Making their statistical analysis we get values of $\mu$ close to the observation made in Section \ref{observation}. Finally we devote Section \ref{end} to illustrating theoretical arguments supporting the main conclusion of this paper that the interaction between faults is a form of criticality identical to that of the human brain.

\section{Observation of California and Hindu-Kush earthquakes} \label{observation}

The statistical analysis of seismic fluctuation is done using the updated version of DEA described in Ref. \cite{garland}. Here we remind the readers about the main properties of the old \cite{scafetta} and new DEA version \cite{garland}.  To make this illustration we adopt the hypothesis done by Helmsmettler and Sornette \cite{subordination}  that seismic processes can described using a generalized form of Continuous Time Random Walk \cite{ctrw}. This is equivalent to assuming \cite{subordination2} that in the operational time scale $n$, the process is described the diffusion equation 
\begin{equation}
\frac{\partial}{\partial n} p(x,n) =  \mathcal{L}p(x,n),
\end{equation} 
where
\begin{equation}
\mathcal{L} \equiv D \frac{\partial^{\beta}}{\partial x^{\beta}},
\end{equation}
with $\beta \leq 2$.  Making subordination with the waiting time distribution density of Eq. (\ref{manneville}) yields \cite{subordination,subordination2}
\begin{equation}
\frac{\partial}{\partial t} p(x,t) = \int_{0}^{t} \Phi(t') \mathcal{L}  p(x,t-t')dt',
\end{equation}
where $\Phi(t)$ is the Montroll-Weiss memory kernel defined by its Laplace Transform
\begin{equation}
\hat \Psi(u) = \frac{u \hat \psi(u)} {1 - \hat \psi(u)}.
\end{equation}
This diffusion process fits the scaling condition $x \propto t^{\delta}$,  where
the scaling power $\delta$ is given by
\begin{equation} \label{abigail}
\delta = \frac{\mu-1}{\beta}. 
\end{equation}

The old generation DEA \cite{scafetta}  evaluates correctly the scaling of  Eq. (\ref{abigail}). This allows us to explain the results reported in Tables 6-9 of the work of Jim\'{e}nez \cite{abigail}, showing  a wide  variety of conditions in the California earthquakes. The DEA scaling $\delta$ undergoes big changes from very small sub-diffusional 
values, $0.183$ to very large super-diffusion, 0.853.  According to Eq. (\ref{abigail}) crucial events with $\mu-1 = 2$ may generated very large scaling if in the operational time scale $\beta$ is very small, namely the intensity of fluctuation is very large. On the other hand if $\mu$ is very close to $1$, the scaling can be very small even if $\beta = 2$.

The results of this paper leads us to believe that $\mu > 2$. Thus the sub-diffusional scaling of Ref. \cite{abigail} is probably the consequence of the joint action of memory and renewal events \cite{subordination2}. This is an interesting issue out of scope of the present paper. We limit ourselves to stressing that the scaling evaluation  in this paper is done with the updated DEA version of Ref. \cite{garland}.  The seismic fluctuations are observed by using the method of stripes. The ordinate axis of the observed fluctuations is divided in bins of size $s$ and an event is defined as the crossing or touching the lines dividing different bins. These events can either be crucial or not crucial. The non-crucial events can be random noise or correlated Gaussian fluctuations. In both cases an ordinary scaling would be generated by this method. In the case $2 < \mu < 3$, the scaling generated by a random walker that at the occurrence of an event makes a step ahead is

\begin{equation}
\delta = \frac{1}{\mu -1}.
\end{equation}

In the long time the DEA scaling depends only on the larger scaling of the crucial events. 

\subsection{California earthquakes} \label{california}

The data analyzed in this section are derived from Ref. \cite{Ncali} and Ref. \cite{caltech}. 

\begin{figure}
	\begin{center}
		\includegraphics[width=1\linewidth]{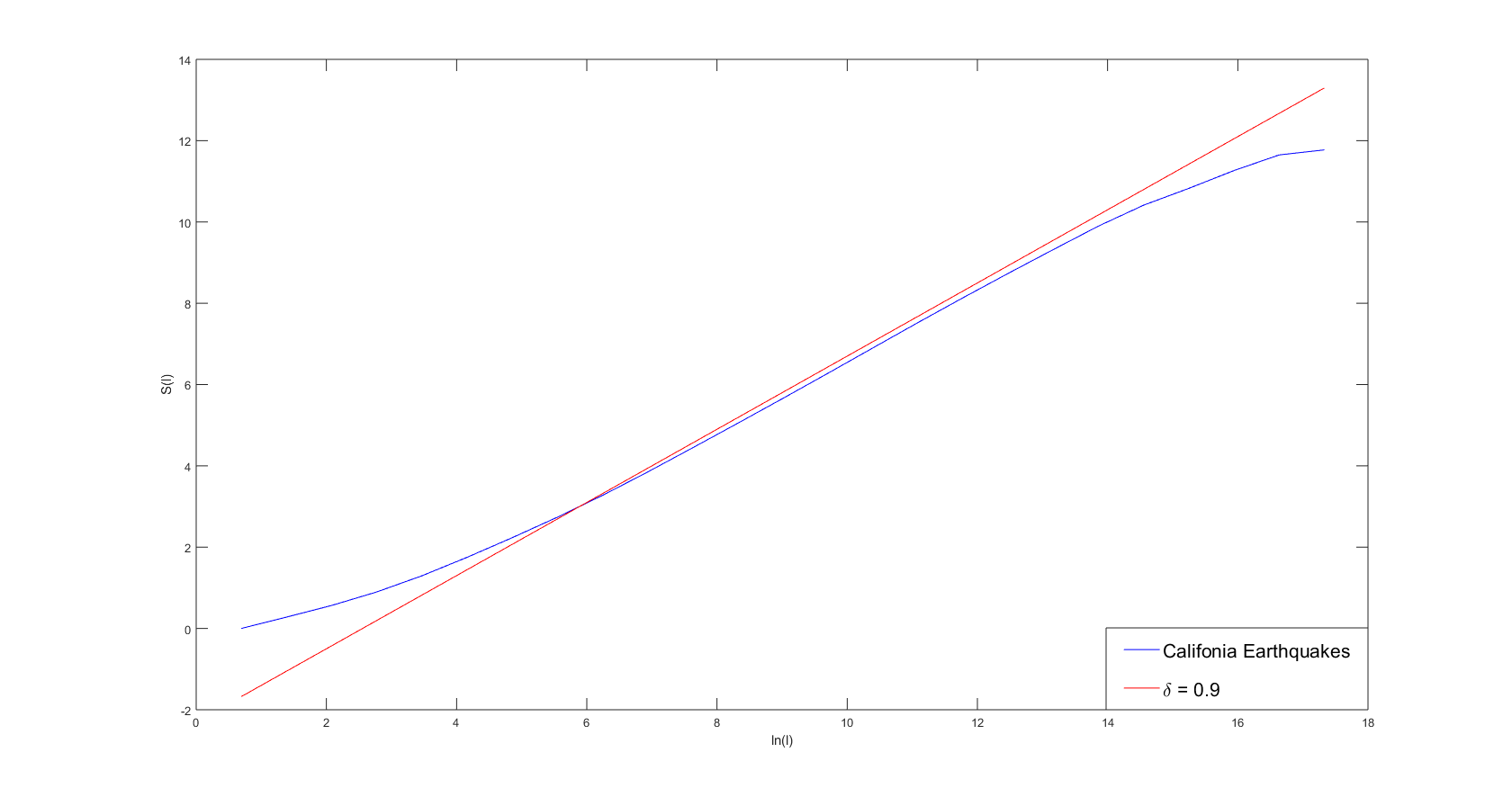}
		\caption{DEA of California Earthquakes.}
		\label{FIG2}
	\end{center}
\end{figure}

California is split by boundary between the Pacific plate and North American plate, which are sliding past one another due to convection currents throughout the Earth’s mantle and crust, generating the source of much
seismic activity in the region. This Transform plate boundary is dominated by the right Lateral lateral strike slip fault in California, known as the San Andreas fault zone. The long-term motion of the two tectonic plates has caused many faults to appear around the plate boundary, where earthquakes of small and
significant magnitudes can occur. These faults include strike-slip faults exhibiting predominantly horizontal motion, but dip-slip (vertical motion), and oblique faults horizontal and vertical motions can arise in the region to account for the overall plate motion in localized areas \cite{faultm}. Data in this region were accessed from the USGS online database \cite{usgs}.
Analyzing this series of earthquakes that occurs throughout California using DEA results in a $\delta = 0.9$, corresponding to $\mu = 2.11$. See Fig. \ref{FIG2}. 

We then take a look at a smaller region of seismic activity, which occurred in July of 2019,  within and around Ridgecrest, California. The Ridgecrest sequence consisting of 2 main shocks, M6.4 and M7.1 36 hours later, comprised of complex surface ruptures but particularly two nearly perpendicular fault planes. This smaller region, compared to the region mentioned earlier,   yields a value of $\mu$ larger than  that generated by a large region composed of numerous faults. This  small region is assumed to have only 2-3  main faults as shown in FIG {\ref{ridgefault}. However, in spite of not being a   single fault, we  assume that they exhibit a behavior closer to that of an individual fault than to the behavior of a large network. Looking at Fig \ref{ridge} we see that this area of interest has a $\mu = 2.39$ which is larger than that of the large area studied earlier, and areas of multiple faults in general.

\begin{figure}
	\begin{center}
		\includegraphics[width=0.8\linewidth]{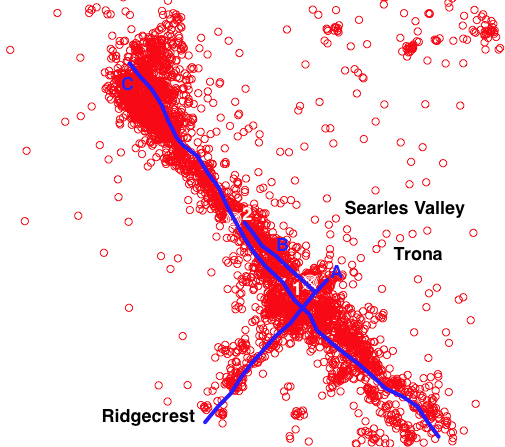}
		\caption{Main fault lines in the Ridgecrest region of interest. The red circles indicate an earthquake epicenter and the blue lines show main fault lines, the location 1 is the 6.4 magnitude epicenter and 2 is the 7.1 magnitude epicenter.}
		\label{ridgefault}
	\end{center}
\end{figure}

\begin{figure}
	\begin{center}
		\includegraphics[width=1\linewidth]{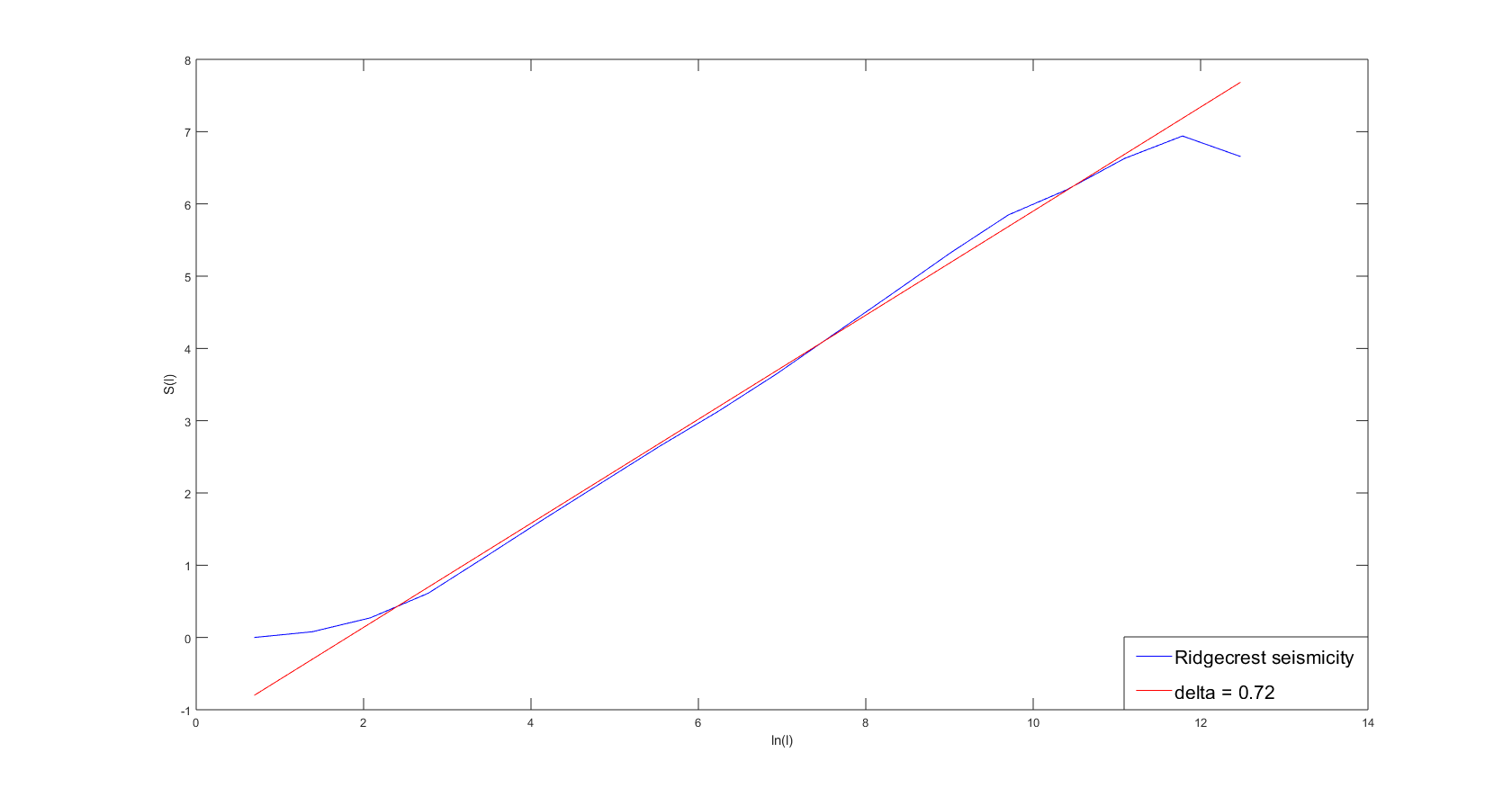}
		\caption{DEA of Earthquakes in Ridgecrest California.}
		\label{ridge}
	\end{center}
\end{figure}

\subsection{Hindu Kush earthquakes}

In the region of Hindu-Kush, stretching from Afghanistan to Pakistan, there are many large magnitude Earthquakes that occur and is considered one of the most seismically active terrains. This terrain is dominated by intermediate-depth earthquakes and mark a poorly defined subduction zone. Differently than the California Region, this Convergent plate boundary consists of tectonic plates colliding into one another. These collisons are responsible for the dip slip and oblique faults in the region but also some strike slip faults in result of the overall compression. We use the complete, homogenized and de-clustered catalog of Rehman et al. \cite{kushdata}.  Analyzing this data using DEA gives a $\delta = 0.91$ as shown in Fig. \ref{FIG3}, corresponding to $\mu = 2.1$ which is nearly identical to the value found in California earthquakes. We make use of geographical data \cite{kushmap} to show that there are 12 significant faults in the region of interest (see Fig \ref{hind}). These faults will have some interaction as plates move past one another and the Earth's surface shifts.

\begin{figure}
	\begin{center}
		\includegraphics[width=1\linewidth]{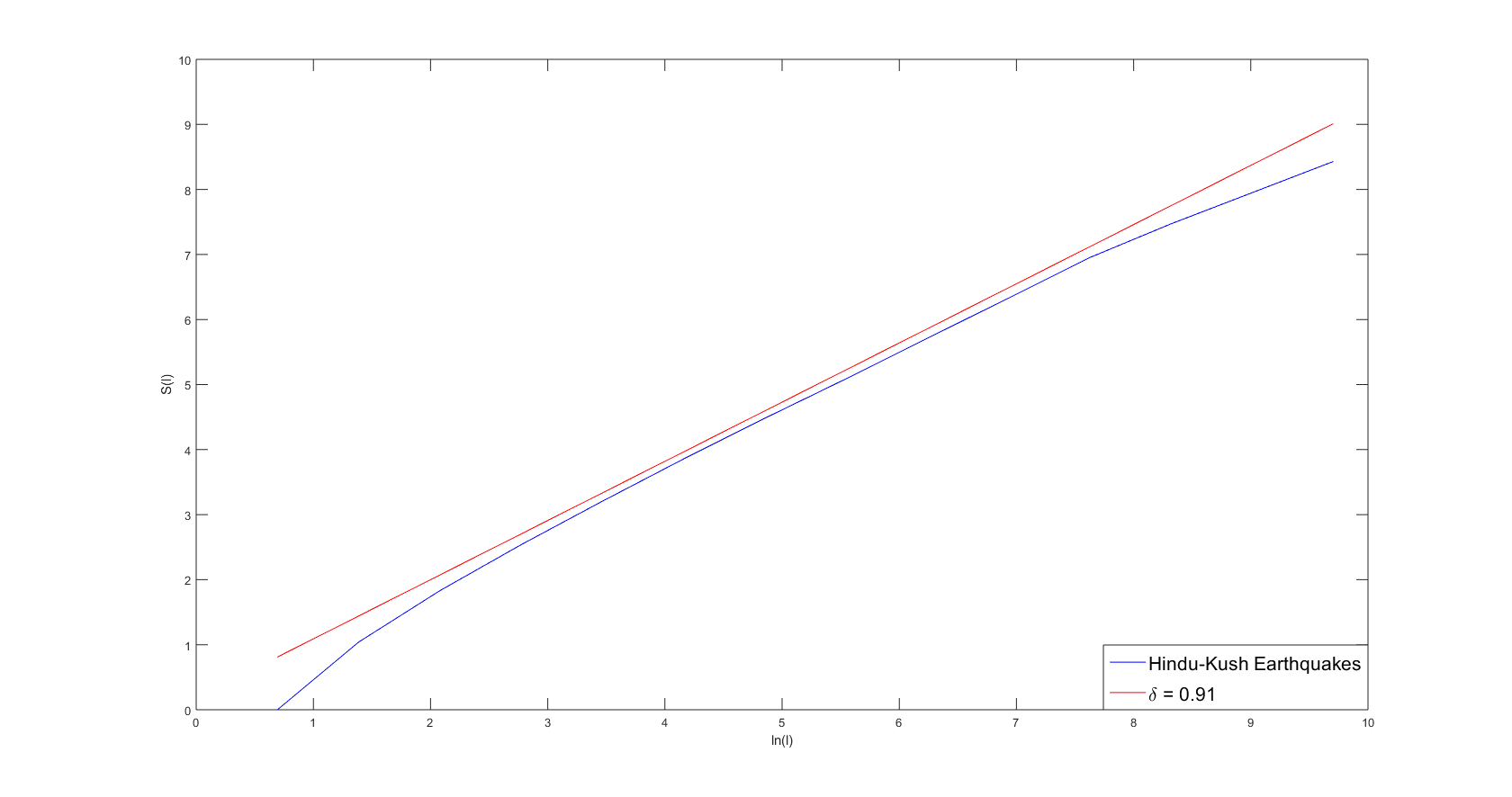}
		\caption{DEA applied to data on Hindu-Kush Earthquakes.}
		\label{FIG3}
	\end{center}
\end{figure}

\begin{figure}
	\begin{center}
		\includegraphics[width=0.9\linewidth]{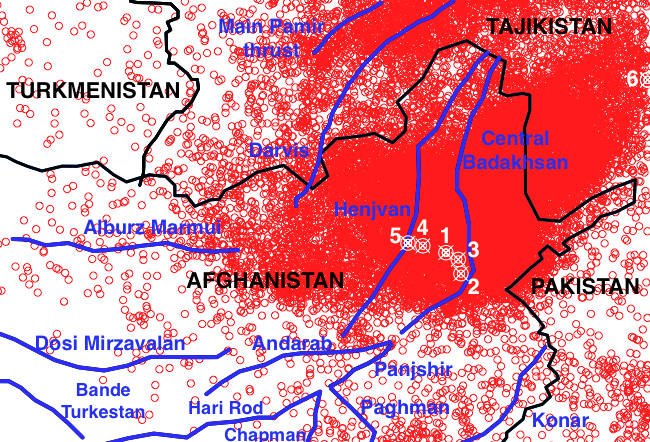}
		\caption{We show the faults in our region studied, labeled by the blue lines. The solid black lines are the country borders and the red circles represent the earthquakes in the region. Six of these earthquakes have a magnitude greater than 6.9 indicated by the white numbers.}
		\label{hind}
	\end{center}
\end{figure}

\section{A model for a single fault} \label{tosatti}
The authors of Ref. \cite{mega}, see their Fig. 1, assumed that the distance between two consecutive main shocks is given by the inverse power law waiting distribution density of Eq. (\ref{manneville}). They claimed that the DEA analysis adopted by them revealed the inverse power law index $\mu$, according to the prescription
\begin{equation}
	\mu = 1 + \frac{1}{\delta},
\end{equation}
where $\delta$ is the scaling detected by DEA.  However, they did not have a model to generate surrogate time sequences of seismic fluctuations. We believe that Braun and Tosatti propose a model corresponding to the intuition of the authors of Ref. \cite{mega}. Therefore in this section we bypass the limitations of the earlier work of Ref. \cite{mega} by creating surrogate sequences based on the model of  Ref. \cite{erio}. 
In this section we apply the updated version of DEA \cite{garland} to surrogate sequence that we create with the model of Ref. \cite{erio}.

We see that the model of Braun and Tosatti \cite{erio} yields $\mu = 2.4$, which significantly larger than $\mu = 2.06$ of Ref. \cite{mega}). On the basis of the results of Section \ref{california} we make the conjecture that this discrepancy is due to the fact in this section we studied a single fault. 

We assume that the spring-block model of \cite{erio} is the realization of a single fault. This assumes the fault is made up of several macrocontacts each connected by springs of constant K. These macrocontacts are also coupled to the ground by a constant k which has a dependence on k, and move across the surface with some velocity v. As they move, there is an increase in the shear force $F_i$ which increases linearly as $kv{\cdot}dt$. There is a threshold force $F_{si}$ that keeps the macrocontacts from breaking, which evolves stochastically over time. These thresholds age and evolve individually in such a way due to continuously breaking and reforming. When the shear force exceeds the threshold force there is a force drop (${\Delta}F_i$) on the contact and it breaks, causing the shear force to return to 0 as well as developing a new threshold $F_{si} \sim 0$. 

For the evolution of the threshold force we take the simple Langevin equation 
\begin{equation}
 \frac{dFsi}{dt} = K(F_{si}) + G\xi(t) 
\end{equation}
with corresponding Fokker-Planck equation for the distribution of thresholds $P_c(F_{si};t)$
\begin{equation}
\frac{\partial P_c}{\partial t} + \frac{dK}{dF_{si}}P_c + {K}{\frac{{\partial P_c}}{\partial F_{si}}} = \frac{G^2}{2}\frac{\partial^2 P_c}{\partial {F_{si}}^2}
\end{equation}

where the drift force \cite{peyard} is given by
\begin{equation}
K(Fsi) = {\frac{2{\pi}F_s}{\tau_0}}\beta^2\frac{1 - \frac{F_{si}}{F_s}}{1 + \epsilon{(\frac{F_{si}}{F_s})}^2}
\end{equation}

and the amplitude of the stochastic force is 

\begin{equation}
G = {(\frac{4\pi}{\tau_0})}^\frac{1}{2}\beta{\delta}F_s
\end{equation}

For the model we choose parameters $F_s = 1$, $\beta = 1$, $\tau_0 = 2\pi$, k = 0.03, $\epsilon = 75$, $v = 0.01$, $dt = 0.01$ and $\delta{F_s} = 0.1F_s$ following the prescription and other initial conditions of \cite{erio}. When a contact breaks, a new threshold is formed it is taken from the distribution with average of $F_{avg} = 0.01F_s$ and deviation of $F_\delta = F_{avg}$. To get the amplitude of the earthquake we sum all force drops of each macrocontact at each time step, given by equation \ref{ampl}, over the total length of the time series. 
\begin{equation} 	\label{ampl}
A = \sum_{i}  \frac{{\vartriangle}F_i(t)}{F_s} 
\end{equation}

Once we have our time series of amplitudes, which represent magnitudes of earthquakes, we can apply DEA. After analyzing our series we see that $\delta = 0.73$ as depicted in Fig. \ref{FIG4}, that corresponds to a $\mu$ of 2.37. This result, from a single fault, exhibits similar results to that of real data analyzed in section \ref{california}. This leads us to believe that this is a plausible model for region studied with a minimal amount of faults as it closely correlates to the data analyzed from the Ridgecrest Earthquakes in section \ref{california}.

\begin{figure}
	\begin{center}
		\includegraphics[width=1\linewidth]{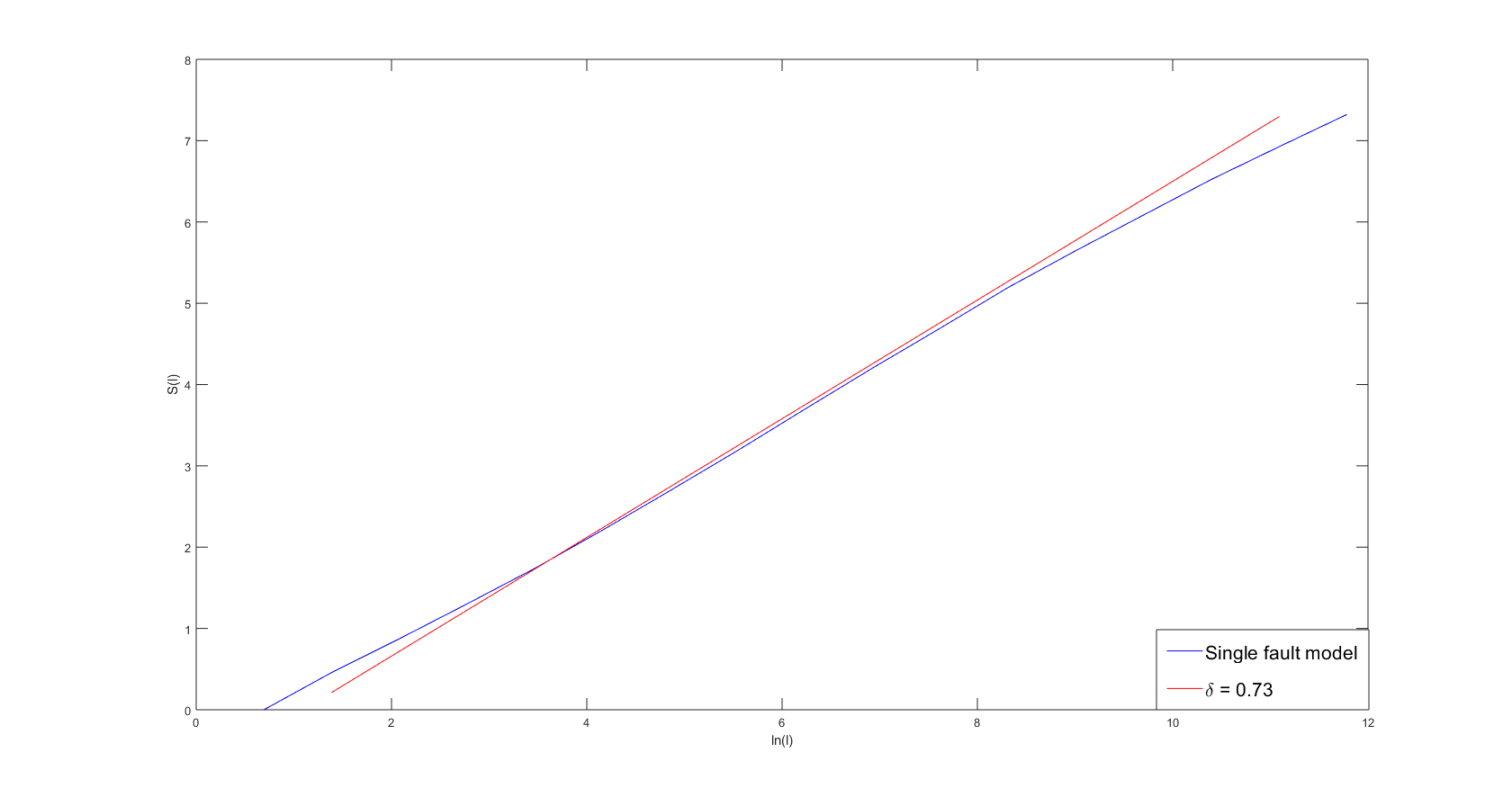}
		\caption{DEA of the single fault model.}
		\label{FIG4}
	\end{center}
\end{figure}

\section{A model for the interaction of faults} \label{interacting}

For the multi-fault (MF) model we imagine a very simple expansion of the single fault model of Section \ref{tosatti}. We assume there are several single faults, made up of a chain of macrocontacts following the prescription of the single fault model. When one of these faults causes a significant earthquake, each other fault becomes perturbed. For the model we pick a reasonable amount of times a significant earthquake could occur and randomly distribute these influences at different unique times using a uniform distribution. Making this assumption we do not need to run each fault simultaneously, only realize that it evolves similarly to each other fault. We also make the assumption that each fault is the same length and has the same number of macrocontacts. We further assume that not all influence is equal based on many factors such as distance and propagation of the wave, hence we adopt a random increase in the shear force to model the influence received from another breaking fault. This compensates for earthquakes that happen at the same time at different faults and reach a contact at the same time. For the simulation we made the assumption that a unit might feel an influence to increase of about 0-30$\%$. In this case when a large Earthquake occurs at another fault, so on the observed fault at the time of the earthquake has shear force $F_i \rightarrow (1.00-1.3)F_i$, then if any $|F_i| > F_{si}$ those contacts will break at that moment. It is possible that no contacts will break at that moment, if a contact feels influence from another fault it will close the gap between the shear force and the threshold force of each contact and can encourage them to break at times different than those that would have been found without this influence. 

For the simulation we used the same parameters as those in Section \ref{tosatti} and assumed a earthquake would happen, on average, around $\frac{7}{100}\%$ of the total time series for a single fault. With this assumption we randomly chose times $\frac{7}{10}\%$ of the total time series to imagine around 10 significant faults in the model as to be on the same order as those of the real earthquakes analyzed. This actual amount of times tends to be slightly lower than that since some earthquakes will trigger at the same time at different faults. Running DEA on the MF model we find that the scaling $\delta \rightarrow 0.86$ as shown in Fig. \ref{FIG5}, corresponding to $\mu = 2.16$. This shift in $\mu$ from the single to MF model is now comparable to that of real earthquakes such as those analyzed in California and the Hindu-Kush region.

\begin{figure}
	\begin{center}
		\includegraphics[width=1\linewidth]{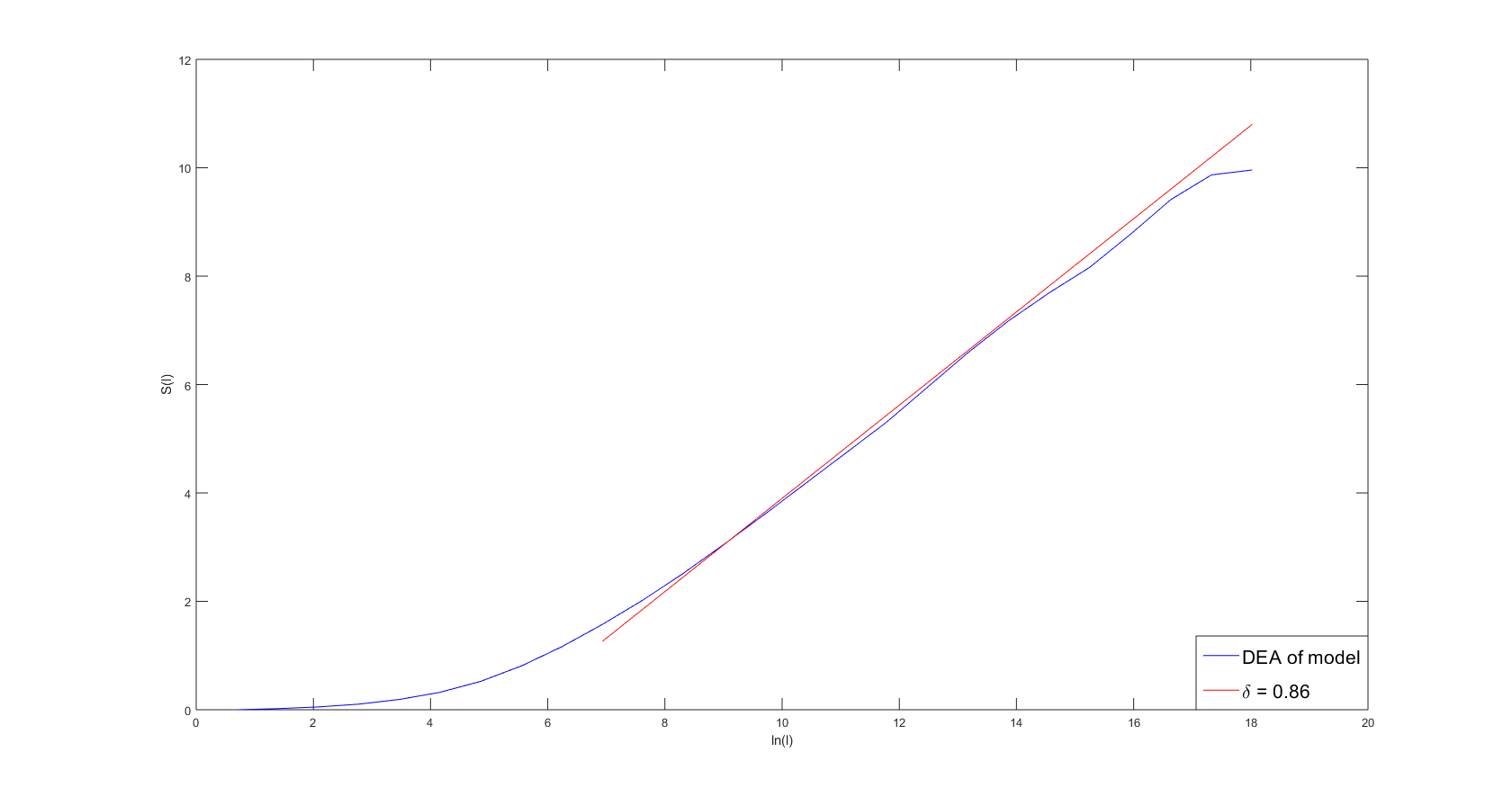}
		\caption{DEA on the interacting fault model.}
		\label{FIG5}
	\end{center}
\end{figure}

\section{conclusions} \label{end}
Through the use of simple spring-block models we show that fault interaction makes the temporal complexity parameter $\mu$ shift from higher values to values closer to $2$, with the value $\mu =2$ corresponding to an ideal $1/f$ noise which was found to characterize the temporal complexity of the brain dynamics in the awake state \cite{brain}. On the basis of earlier work on neurophysiological  and social models, we make the conjecture that this phenomenon is the result of a cooperative interaction between the units of a self-organizing system. The interacting units are different faults.

The communication between faults is still an open research problem \cite{procaccia} requiring further progress to explain the communication between distant faults, and especially the self-amplification of small perturbation. We notice that the boundary conditions are properties of tectonic plates. In California the plates slide passing each other \cite{faultm}. In Hindu Kush the pates collide with one another \cite{kushmap}. In spite of a different way of generating earthquakes, as proved by our DEA analysis, the properties of the temporal complexity indexes $\mu$ are the same, $\mu$ closer to $2$ for the multi-fault condition. The results of  the heuristic model of fault interaction used in this paper  emphasize that the critical nature of earthquakes is independent of fault type and of the action of  either vertical or horizontal motion. While we make note of the fact that the true number of faults is still difficult to asses, we can use the main faults to identify the main units that interact at a significant level. 

For these reasons we would like to finally speculate that an improvement of the model may lead to an even closer agreement between the modeling and the result of DEA applied to wide seismic regions and leading to $\mu = 2.06$ \cite{mega}.  We notice that the DEA analysis applied to bio-photon emission triggered  by the germination seeds yield  $\mu \approx 2.2$, suggesting a kind of complexity matching between the temporal complexity of the germinating seeds and the geophysical background. 
Moving to cognition \cite{eigen}, this is equivalent to realize that establishing a boundary between living and non living matter is an extremely hard problem.

\emph{acknowledgments} Research work done with the support of ARO grant W911NF1901. Seismological background was provided by David Oglesby and seismicity with fault lines were plotted in programming language R with the help from Keith Richards-Dinger. Kuntal Chaudhuri's fitted fault planes of Ridgecrest aftershocks contributed to the fault zone.  Through the efforts of all the assistance, a big thank you is well deserved.

\bibliography{reference}

\end{document}